\documentclass[twocolumn, pra]{revtex4-1}
\usepackage{graphicx}
\usepackage{amsmath}
\usepackage{amssymb}
\usepackage[autostyle]{csquotes}
\usepackage{dcolumn}
\usepackage{float}
\usepackage{bm}
\usepackage[breaklinks=true,colorlinks,citecolor=blue,linkcolor=blue,urlcolor=blue]{hyperref}
\usepackage{enumitem}

\newlist{inlineroman}{enumerate*}{1}
\setlist[inlineroman]{itemjoin*={{, and }},afterlabel=~,label=\roman*.}

\newcommand{\inlinerom}[1]{
\begin{inlineroman}
#1
\end{inlineroman}
}

\DeclareMathAlphabet{\bi}{OML}{cmm}{b}{it}
\def\be{\begin{equation}}
\def\ee{\end{equation}}
\def\bearr{\begin{eqnarray}}
\def\eearr{\end{eqnarray}}
\def\la{\langle}
\def\ra{\rangle}

\begin{document}
\title{The Levy-Perdew-Sahni equation for the density: accurate effective potential and self-consistent solution for two-electron atomic systems}
\bigskip
\author{Rabeet Singh Chauhan }
\email[E-mail: ]{rabeet@iitk.ac.in}
\author{Manoj K. Harbola}
\email[E-mail: ]{mkh@iitk.ac.in}
\normalsize
\affiliation
{Department of Physics, Indian Institute of Technology-Kanpur,
Kanpur-208 016, India}
\date{\today}
\begin{abstract}
 Levy-Perdew-Sahni (LPS) provided a proof of the connection between the density and ionization potential of a many-electron system using the equation for the density directly. This equation employs an effective potential which is derived by LPS from the many-electron wavefunction of the system. In this paper, we calculate this potential explicitly by employing an accurate wavefunction for two-electron systems. With this wavefunction connection the LPS equation is then solved self-consistently employing this potential and shown to lead to accurate densities and chemical potentials for these systems.  
\end{abstract}
\maketitle
\section{Introduction}
The asymptotic behavior of the ground state electron density $\rho(\vec{r})$ of a many-electron system in the limit of $|\vec{r}|\rightarrow \infty$  is determined by its ionization potential $I$ and is given as (atomic units are used throughout the paper)
\begin{equation}
\rho(\vec{r},|\vec{r}|\rightarrow \infty) \propto e^{-2 \sqrt{2 I} r} \label{eq1}
\end{equation}
This has been shown in many different ways. Some of these are as follows : Morrel, Levy and Parr \cite{MPL, PL} employed natural spin orbitals formed from the many-electron wavefunction to prove this. Katriel and Davidson \cite{Katr} derived the asymptotic form by analytic continuation of asymptotic scattering-state wavefunctions.  von-Barth \cite{Barth} worked in terms of quasi-particle amplitudes to show the above form. All these investigations were carried out on the basis of many-electron wavefunction. In contrast Levy-Perdew-Sahni \cite{LPS} employed the density directly and proved Eq. (\ref{eq1}). To do this they derived and used the following equation (referred to as LPS equation) for the density  
\begin{equation}
[-\frac{1}{2}\nabla^2 + v_{ext}(\vec{r}) + v_{eff}(\vec{r})]\rho^{\frac{1}{2}}(\vec{r}) = \mu  \rho^{\frac{1}{2}}(\vec{r}). \label{eq2}
\end{equation}
In this equation $v_{ext}(\vec{r})$ is the external potential $-$ like the nuclear potential in an atom $-$ that electrons are moving in and $\mu$ is the chemical potential. The latter is defined \cite{PPLB} to be equal to the negative of the ionization potential. Furthermore, $v_{eff}(\vec{r})=v_{H}(\vec{r})+v_{xc}(\vec{r})+v_{KE}(\vec{r})$ is the effective potential that is the sum of the Hartree potential \cite{PARR},
\begin{equation*}
v_{H}(\vec{r})=\int \frac{\rho(\vec{r'})}{|\vec{r}-\vec{r'}|} d \vec{r'},
\end{equation*} 
the exchange-correlation potential \cite{PARR}
\begin{equation}
v_{xc}(\vec{r})=\frac{\delta E_{xc}[\rho]}{\delta \rho(\vec{r})} \label{eq3}
\end{equation}
where $E_{xc}[\rho]$ is the unknown exchange-correlation energy functional and a kinetic energy component 
\begin{equation}
v_{KE}(\vec{r})= \frac{\delta \tilde{T_s}[\rho]}{\delta \rho(\vec{r})} \label{eq4}
\end{equation}
where
\begin{equation}
\tilde{T_s}[\rho] = T_s[\rho]-\la \rho^{\frac{1}{2}}|-\frac{1}{2}\nabla^2|\rho^{\frac{1}{2}} \ra \label{eq5}
\end{equation}
is obtained by subtracting $\la \rho^{\frac{1}{2}}|-\frac{1}{2}\nabla^2|\rho^{\frac{1}{2}} \ra$ from the non-interacting kinetic energy functional $T_s[\rho]$ for a given ground state density. This term is also referred to as the Pauli-term with $v_{KE}(\vec{r})$ \cite{Nagy, March} as the Pauli-potential. It is evident that for a two-electron system $\tilde{T_s}[\rho]$ vanishes and $v_{eff}(\vec{r})$ is the sum of the Hartree and the exchange-correlation potential. Since the functionals $E_{xc}[\rho]$ and $T_s[\rho]$ are not known, $v_{eff}(\vec{r})$ of the LPS equation cannot be calculated exactly using equations (\ref{eq3}) and (\ref{eq4}). We note though that the exchange-correlation potential can be constructed \cite{Zhao, Van} for a given ground state density.  Similarly $v_{eff}(\vec{r})$ can be calculated as
\begin{equation*}
v_{eff}(\vec{r})=\frac{1}{2}\frac{\nabla^2 \rho^{\frac{1}{2}}}{\rho^{\frac{1}{2}}} - v_{ext}(\vec{r})+\mu.
\end{equation*}
However this equation cannot be employed to gain insights into the behavior of $v_{eff}(\vec{r})$ that are needed to get the asymptotic structure of the density. To study the latter, LPS derived an expression for $v_{eff}(\vec{r})$ in terms of the many-electron wavefunction of the system and used this to prove that if for $|\vec{r}|\rightarrow \infty$, the density decreases as  
\begin{equation}
\rho(\vec{r}) \rightarrow e^{-2 \alpha r} \label{eq6}
\end{equation}
then 
\begin{equation}
\alpha \geq \sqrt{-2 \mu}. \label{eq7}
\end{equation}
On the other hand, using Eq. (\ref{eq2}) it can also be shown \cite{LPS}
\begin{equation}
\alpha \leq \sqrt{-2 \mu} \label{eq8}
\end{equation}
under certain conditions. From Eq. (\ref{eq7}) and Eq. (\ref{eq8}) it follows that
\begin{equation}
\alpha = \sqrt{-2 \mu}. \label{eq9}
\end{equation}
It is clear that expression of $v_{eff}(\vec{r})$ in terms of the many electron wavefunction is crucial in developing an understanding of the behavior of $v_{eff}(\vec{r})$ and hence in proving Eq. (\ref{eq1}).
\par
Although $v_{eff}(\vec{r})$ in terms of the many-electron wavefunction is known, it has never been calculated explicitly. The reason for this is obvious : since $v_{eff}(\vec{r})$ is written in terms of the exact eigenfunction, its calculation requires highly accurate wavefunctions.  Thus even for two-electron system where the potential $v_{eff}(\vec{r})$ is known to be the sum of the Hartree and the exchange-correlation potential, such a direct calculations has not been performed. As a result, the LPS equation for the density has never been solved exactly. This paper fills this gap for two-electron systems. In the following we first briefly review the main features of the LPS equation and reproduces the expression for $v_{eff}(\vec{r})$ in terms of the many-electron wavefunction. We then describe the wavefunction that we employ to calculate $v_{eff}(\vec{r})$ for two-electron atoms and use it to construct $v_{eff}(\vec{r})$. We solve the LPS equation employing the $v_{eff}(\vec{r})$ so obtained and show that the resulting densities and ionization potentials are in excellent agreement with those calculated from the corresponding wavefunction.
\par 
Having solved the equation for $v_{eff}(\vec{r})$ constructed from an accurate wavefunction, we next perform a self-consistent calculation using the LPS equation for the ground state density of two-electron atoms. We can do this because of the form of correlated wavefunction employed by us. This allows us to write the correlated wavefunction in terms of $\rho^{\frac{1}{2}}(\vec{r})$ and calculate $v_{eff}(\vec{r})$ from it. We show that the solution converges to highly accurate results for the density and the chemical potential.
\section{The Levy-Perdew-Sahni equation}
The equation derived by LPS for the density is that given in equation (\ref{eq2}). A time-dependent version of this equation was proposed independently by Chattaraj and Deb \cite{BMD} and solved by calculating $v_{eff}(\vec{r})$ employing approximate exchange-correlation and kinetic energy functional. Recently the equation has also been solved for the $Be$ atom by developing an equation for the Pauli-potential given by Eq. (\ref{eq4}) and using an approximate energy functional for the exchange and correlation potential, thus still calculating $v_{eff}(\vec{r})$ approximately \cite{Nagy}. LPS, however, gave an exact expression for $v_{eff}(\vec{r})$ in terms of the many-electron wavefunction $\Psi(\vec{x}_1,  \vec{x}_2,..\vec{x}_N)$ for $N$ electrons. This is given as 
\begin{widetext}
\begin{equation}
 v_{eff}(\vec{r}_{N}) = \int \frac{\tilde{\rho}_{N-1}(\vec{r};\vec{r}_N)}{|\vec{r}_N-\vec{r}|} d\vec{r}+ \la \phi_{N-1} |H(N-1)-E_{N-1}^0| \phi_{N-1} \ra + \frac{1}{2}\int \vec{\nabla}\phi_{N-1} \cdot \vec{\nabla} \phi_{N-1} d\vec{x}_1 d\vec{x}_2 ...d\vec{x}_{N-1} d\sigma_N \label{eq10}
\end{equation}
\end{widetext}
where $x$ represents both the space variable $\vec{r}$ and spin variable $\sigma$.  Further, $\phi_{N-1}(\vec{x}_1,  \vec{x}_2, \vec{x}_3, ...\vec{x}_{N-1}; \vec{r}_N,\sigma_N)$ is a function defined via the equation 
\begin{equation}
\Psi(\vec{x}_1,  \vec{x}_2,..\vec{x}_N)=\frac{1}{\sqrt{N}}\rho^{\frac{1}{2}}(\vec{r}_N) \phi_{N-1}(\vec{x}_1,  \vec{x}_2,...\vec{x}_{N-1}; \vec{r}_N,\sigma_N) \label{eq11}
\end{equation}
where $\Psi$, is the many-electron wavefunction.  $H(N-1)$ is defined by splitting $H(\overline{N})$ as 
\begin{equation} 
H(N)=-\frac{1}{2} \nabla^{2}_{\vec{r}_N} + v_{ext}(\vec{r}_N) + H(N-1) +  \sum_{i=1}^{N-1} \frac{1}{|\vec{r}_i-\vec{r}_N|}.\label{eq12}
\end{equation}
and $E_{N-1}^0$ is the ground-state energy of the $N-1$ electron system. Note that $\phi_{N-1}(\vec{x}_1,  \vec{x}_2, \vec{x}_3, ...\vec{x}_{N-1}; \vec{r}_N,\sigma_N)$ depends parametrically on $\vec{r}_N$ and $\sigma_N$. Thus $\vec{\nabla_N}\phi_{N-1}(\vec{x}_1,  \vec{x}_2, \vec{x}_3, ...\vec{x}_{N-1}; \vec{r}_N,\sigma_N)$ is not zero in general. Finally $\tilde{\rho}_{N-1}(\vec{r}; \vec{r}_N)$ is the density corresponding to $\phi_{N-1}$ with parametric dependence on $\vec{r}_N$. In the expression for $v_{eff}(\vec{r})$, each term is independently greater than zero; the first and the third term are so because of the positive integrand and the second term is positive by the variational principal for the energy. Thus $v_{eff}(\vec{r}) \geq 0$ for all values of $\vec{r}$. If  for $\vec{r} \rightarrow \infty$, $\rho(r)$ decays as $ e^{- 2 \alpha r}$ then from equation (\ref{eq2}) 
\begin{equation}
\alpha = \sqrt{-2 (\mu-v_{eff}(\infty))}. \label{eq13}
\end{equation}
With $v_{eff}(\vec{r}) \geq 0$,  equation (\ref{eq7}) i.e. $\alpha \geq \sqrt{-2\mu}$ follows and the equality holds if $v_{eff}(\infty)=0$. The potential $v_{eff}(\infty)$ vanishes as $|\vec{r}|\rightarrow \infty$ if $\phi_{N-1}$ collapses to the exact ground state function $\Psi_{N-1}(\vec{x}_1,  \vec{x}_2, \vec{x}_3, ...\vec{x}_{N-1})$ of the corresponding ion. To get the equality $\alpha = \sqrt{-2 \mu}$, LPS further consider the equation for the function 
\begin{equation}
g_l(\vec{r}_N, \sigma_N)= \rho^{\frac{1}{2}}(\vec{r}_N)\la \Psi_{N-1}^l|\phi_{N-1} \ra  \label{eq14}
\end{equation} 
and used this to prove Eq. (\ref{eq8}) i.e. $\alpha \leq \sqrt{-2\mu}$. This can be consistent with $\alpha \geq \sqrt{-2 \mu}$ only if $\alpha=\sqrt{-2\mu}$.
\par
It is clear from the description above that structure and properties of $v_{eff}(\vec{r})$ are the basis of the proof by LPS. As stated earlier, the potential $v_{eff}(\vec{r})$ has never been obtained in its exact form starting from a wavefunction even for two-electron systems where $v_{eff}(\vec{r})$ has no kinetic energy component and is thus equal to the sum of the Hartree and the exchange-correlation potential. In this paper we calculate $v_{eff}(\vec{r})$ starting from an accurate wavefunction for 2-electron atomic systems. After calculated the effective potential we solve the LPS equation both \inlinerom{\item for $v_{eff}(\vec{r})$ calculated from the accurate wavefunction employed by us \item self-consistently.}We show that the solution gives  highly accurate densities and ionization energies for the these systems.
\par 
For a two-electron atomic system, the electronic wavefunction $\Psi_2(\vec{r}_1, \vec{r}_2)$ is a singlet with the electrons having opposite spins. Thus
\begin{equation}
\phi_1(\vec{r}_1; \vec{r}_2) = \sqrt{\frac{2}{\rho(\vec{r}_2)}}\Psi_2(\vec{r}_1, \vec{r}_2) \label{eq15}
\end{equation} 
where
\begin{equation}
\rho(\vec{r}_2) = 2 \int |\Psi_2(\vec{r}_1, \vec{r}_2)|^2 d\vec{r}_1. \label{eq16}
\end{equation}
Furthermore, 
\begin{equation}
\tilde{\rho}_1(\vec{r}_1; \vec{r}_2) = |\phi_1(\vec{r}_1; \vec{r}_2)|^2 \label{eq17}
\end{equation}
and 
\begin{widetext}
\begin{equation}
v_{eff}(\vec{r}_2)= \int \frac{\tilde{\rho}_1(\vec{r}_1; \vec{r}_2)}{|\vec{r}_1-\vec{r}_2|} d \vec{r}_1 + \la \phi_1 |-\frac{1}{2}\nabla_1^2-\frac{Z}{r_1}+\frac{Z^2}{2}| \phi_1 \ra + \frac{1}{2} \int |\vec{\nabla}_2 \tilde{\rho}_1(\vec{r}_1; \vec{r}_2)|^2 d \vec{r}_1. \label{eq18}
\end{equation}
\end{widetext}
Note the parametric dependence of $\phi_1$ on $\vec{r}_2$ in the equation above. Looking the equation (\ref{eq18}), it is not obvious that $v_{eff}(\vec{r}_2)$ will be equal to the sum of the Hartree and the exchange-correlation potential. In the following we show numerically that it is indeed so employing two different semi-analytic wavefunction. These wavefunctions are described in the next section.
\section{Le Sech and modified Le Sech wavefunction}
\begin{figure}
\caption{Exchange-correlation potentials $v_{xc}^{L}(\vec{r})$ and $v_{xc}^{ML}(\vec{r})$ calculated for the densities obtained from Le Sech and modified Le Sech respectively. These are compared with exchange-correlation potential obtained from the exact density \cite{UG}. \label{fig1}}
\includegraphics[width=0.5\textwidth]{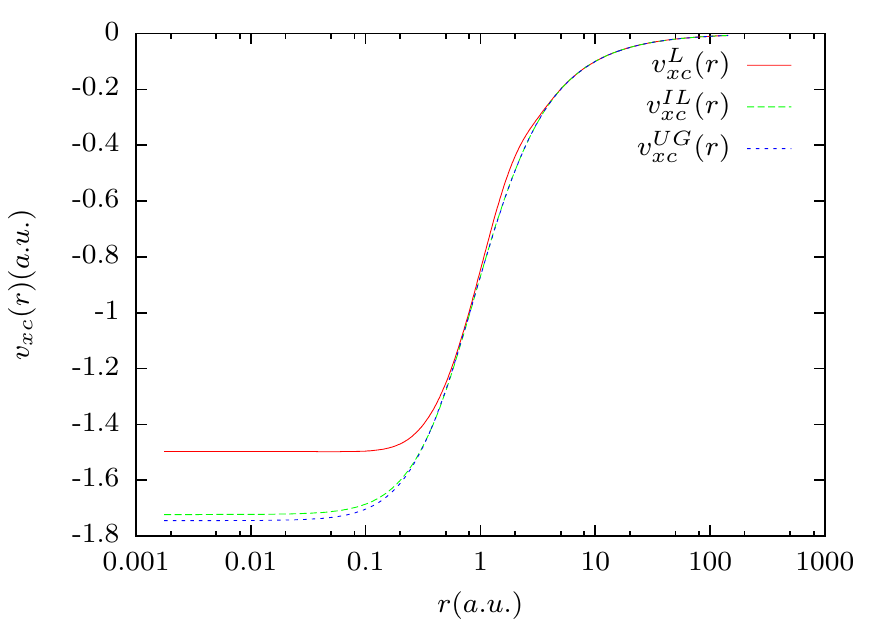}
\end{figure}

To perform fundamental studies in density functional theory, one needs wavefunctions that not only give accurate energies but also equally accurate densities. Furthermore, the wavefunction should have a form that is analytically simple so that it is easily employed for such studies. One such wavefunction is the Le Sech wavefunction \cite{Sech} for two-electron atoms. It is given as 
\begin{equation}
\Psi_L(\vec{r}_1, \vec{r}_2)=C_N e^{-Z r_1} e^{-Z r_2} f(r_1, r_2, r_{12}) \label{psil}
\end{equation}
with
\begin{equation}
f(r_1, r_2, r_{12}) = (\cosh ar_1 + \cosh ar_2)(1+0.5 r_{12} e^{-b r_{12}}) \label{eq20}
\end{equation}
where $C_N$ is the normalization constant and $a$ and $b$ are the variational parameters. The wavefunction leads to energies that are within $0.1\%$ of the exact ground state energies\cite{Sech} for these systems. On the other hand the density that this wavefunction gives is not as accurate as the energies. For example, if the exchange-correlation potential $v_{xc}^L(\vec{r})$  is calculated for this density (we do it using the Zhao-Parr method \cite{Zhao}), it differs from the exact exchange-correlation potential near the nucleus significantly. This is shown in Fig. (\ref{fig1}) where we have plotted the exchange-correlation potentials $v_{xc}^{L}(\vec{r})$ for the He atom and compared it to the exchange-correlation potential $v_{xc}^{UG}(\vec{r})$ that obtained from the accurate density of Umrigar and Gonze \cite{UG} (UG). It is evident that from the nucleus upto about $0.2$au, the exchange-correlation potential obtained from the Le Sech density is less deep with the difference being close to $15\%$. 
\par
The wavefunction can be made more accurate by writing it in the modified form (Modified Le Sech wavefunction \cite{Rabi, Baber}) as 
\begin{equation}
\Psi_{ML}(\vec{r}_1,\vec{r}_2)= \phi (\vec{r}_1) \phi (\vec{r}_2) f(r_1, r_2, r_{12}) \label{psiml}
\end{equation}     
where for each $a$ and $b$, $\phi(\vec{r})$ is obtained by solving a Schr$\ddot{o}$dinger-like equation for it. The function $\phi(\vec{r})$ is suitably normalized so that $\int |\Psi_{ML}(\vec{r}_1, \vec{r}_2)|^2 d\vec{r}_1 d\vec{r}_2=1$. The appropriate ground state wavefunction corresponds to those $a$ and $b$ that lead to the least value of the energy. The optimized $\Psi_{ML}(\vec{r}_1, \vec{r}_2)$ thus obtained improves \cite{Rabi} both the energies and the densities  for two electron systems. This is evident from Fig. (\ref{fig1}) where we have also plotted the exchange-correlation potential $v_{xc}^{ML}(\vec{r})$ for this density obtained from the modified Le Sech wavefunction. It is very close to the $v_{xc}^{UG}(\vec{r})$.
\par
In this paper we calculate $v_{eff}(\vec{r})$ and the corresponding exchange-correlation potential by employing both the Le Sech and the modified Le Sech wavefunctions of Eqs. (\ref{psil}) and (\ref{psiml}) respectively. After constructing $v_{eff}(\vec{r})$, we solve the LPS equation (Eq. \ref{eq2}) for the density by using the $v_{eff}(\vec{r})$ obtained from $\Psi_L(\vec{r}_1, \vec{r}_2)$ and $\Psi_{ML}(\vec{r}_1, \vec{r}_2)$. The solution gives both the density and the chemical potential for the system studied. Finally by writing $\phi(\vec{r})$ of Eq. (\ref{psiml}) in terms of $\rho^{\frac{1}{2}}(\vec{r})$, we also solve the LPS equation self-consistently.
\section{Results}
\subsection{$v_{eff}(\vec{r})$ from accurate wavefunctions}
\begin{figure}
\caption{\label{fig2} Effective potentials ($v_{eff}^{L}(r)$ and $v_{eff}^{ML}(r)$) are constructed using Eq. (\ref{eq18}) for  Le Sech and improved Le Sech wavefunction and $v_{eff}^{SC}(r)$ is calculated by employing the self-consistent solution of the LPS equation.}
\includegraphics[width=0.5\textwidth]{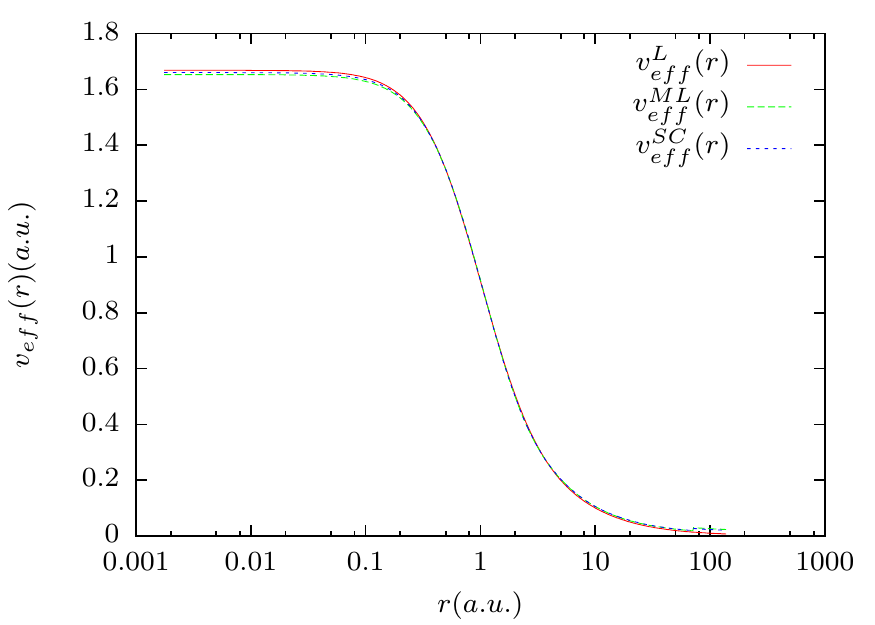}
\end{figure}

\begin{figure}
\caption{\label{fig3} Exchange-correlation potentials are obtained by subtracting the Hatree potential from the $v_{eff}(r)$. $v_{xc}^{SC}(r)$ is obtained from the self-consistent solution of the LPS equation.}
\includegraphics[width=0.5\textwidth]{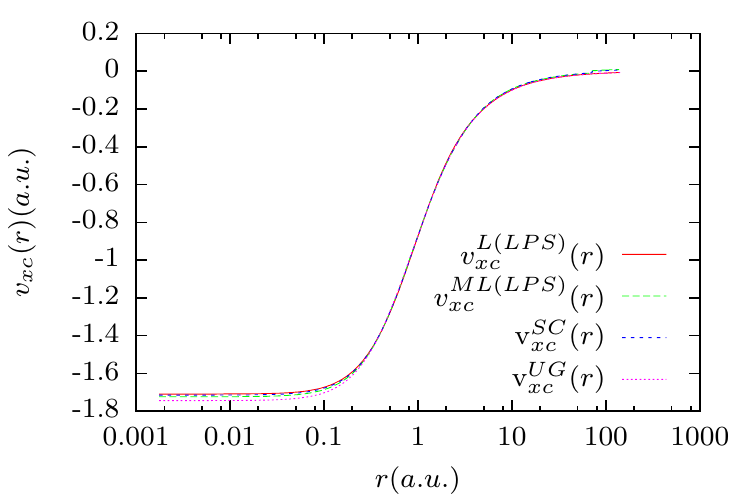}
\end{figure}

Shown in Fig. (\ref{fig2}) is the LPS effective potential $v_{eff}^L(\vec{r})$ calculated for the He atom from the Le Sech wavefunction of Eq. (\ref{psil}). The parameters $a$ and $b$ that we used are those calculated earlier \cite{Rabi}. These values are $a=0.72$ and $b=0.17$. The accuracy of  $v_{eff}^L(\vec{r})$ can be judged by comparing the exchange-correlation potential $v_{xc}^{L(LPS)}(\vec{r})= v_{eff}^L(\vec{r})-v_{H}(\vec{r})$ obtained from it. We show this potential $v_{xc}^{L(LPS)}(\vec{r})$ in Fig. (\ref{fig3}) and compare it with exact the $v_{xc}^{UG}(\vec{r})$ \cite{UG}. We see that the $v_{xc}^{L(LPS)}(\vec{r})$ obtained from $v_{eff}^L(\vec{r})$ is accurate. Note that $v_{xc}^{L(LPS)}(\vec{r})$ is different from the $v_{xc}^L(\vec{r})$ that was calculated from the corresponding density. The difference arises because $v_{xc}^L(\vec{r})$ is that potential that reproduces the density given by the Le Sech wavefunction.  On the other hand, $v_{xc}^{L(LPS)}(\vec{r})$ is calculated directly from the wavefunction. 


\par
The calculation above is also repeated by using modified Le Sech wavefunction of Eq. (\ref{psiml}) with parameters $a=0.93$ and $b=0.20$ \cite{Rabi} and $\phi(\vec{r})$ is obtained by solving the resulting equation. The results are shown in Figs. (\ref{fig2}) and (\ref{fig3}). We see that with the improvement in the wavefunction the exchange-correlation potential (shown as $v_{xc}^{ML(LPS)}(\vec{r})$ in Fig. (\ref{fig3})) also improves, as expected.  
\par
In Table \ref{tb1}, we display the chemical potential obtained as the eigenvalue of the LPS equation solved with the effective potential constructed from the Le Sech and modified Le Sech wavefunctions, respectively, for two-electron systems. These are compared with the energy difference $E_2-E_1$ obtained earlier \cite{Rabi}, where $E_2$ is the energy of the two electron system and $E_1=-\frac{Z^2}{2}$ is the energy of the corresponding ion. Comparison is also made with the experimental chemical potential\cite{Lide} which is equal to negative of the ionization potential. It is seen that the eigenvalues are very close to $E_2-E_1$ as well as experimental chemical potentials.
\begingroup
\squeezetable
\begin{table}[h]
\caption{Chemical potential $\mu$ obtained by solving the LPS equation for $v_{eff}(\vec{r})$ constructed from the Le Sech and modified Le Sech wavefunction. Comparison with the energy difference $E_2-E_1$ is also made for each wavefunction. The experimental chemical potential $=- (I_{exp})$ is given in the last column.\label{tb1}}
\begin{ruledtabular}
\begin{tabular}{lccccc}
 atom &  \multicolumn{2}{c}{Le Sech wavefunction}  & \multicolumn{2}{c}{Modified Le Sech wavefunction} & $I_{exp}$\cite{Lide} \\  \hline
       &  $-\mu$   &$E_1-E_2$& $-\mu$   &$E_1-E_2$& \\         \hline
 H\textsuperscript{$-$} & 0.0267 & 0.0267 & 0.0267 & 0.0271 & 0.0277            \\
 He                     & 0.9038 & 0.9020 & 0.9042 & 0.9028 & 0.9036            \\
 Li\textsuperscript{+}  & 2.7798 & 2.7778 & 2.7807 & 2.7787 & 2.7798            \\
 Be\textsuperscript{2+} & 5.6556 & 5.6533 & 5.6563 & 5.6543 & 5.6557            \\
 B\textsuperscript{3+}  & 9.5311 & 9.5286 & 9.5318 & 9.5296 & 9.5320            \\
 C\textsuperscript{4+}  & 14.4064&14.4038 & 14.4073&14.4050 & 14.4086           \\
 N\textsuperscript{5+}  & 20.2817&20.2789 & 20.2823&20.2801 & 20.2878           \\
 O\textsuperscript{6+}  & 27.1569&27.1541 & 27.1575&27.1552 & 27.1678           \\
 F\textsuperscript{7+}  & 35.0320&35.0291 & 35.0328&35.0303 & 35.0548           \\
 Ne\textsuperscript{8+} & 43.9072&43.9042 & 43.9079&43.9054 & 43.9449 
\end{tabular}
\end{ruledtabular}
\end{table}   
\subsection{Self-consistent solution}
The LPS equation has hitherto not been solved exactly for the density of any system. The usefulness of equation or the effective potential has been to understand the asymptotic decay of the electronic density of  a many-electron system. As discussed in the introduction, this is done by constructing an effective potential from the wavefunction. It was indeed stated in ref. \cite{LPS} that \enquote{The wavefunction connection enables us to understand and prove rigorously various aspects of $v_{eff}(\vec{r})$}.  We now use this connection to solve the LPS equation directly for the square root of density for two-electron atomic systems. These calculations are the first such accurate application of the LPS equation.
\par 
Our calculations proceed as follows : we take the modified Le Sech wavefunction form given by Eq. (\ref{psiml}) with parameters $a$, $b$ and an approximate $\phi(\vec{r})$. From this we calculate the effective potential $v_{eff}(\vec{r})$ and solve the resulting LPS equation for $\rho^{\frac{1}{2}}(\vec{r})$. We then construct the new $\phi(\vec{r})$ for the $i^{th}$ iteration as 
\begin{equation}
\phi_i(\vec{r})=\sqrt{\frac{\rho_i(\vec{r})}{2}} \frac{1}{\sqrt{\int |\phi_{i-1}(\vec{r}_2)f(r_{12})|^2 d \vec{r}_2}} \label{eq21}
\end{equation}  
where $\rho_i(\vec{r})$ is the density obtained from solving the LPS equation in $(i-1)^{th}$ iteration. Thus for a given $a$ and $b$, starting from an approximate $\phi_{i=0}(\vec{r})$ we get new density $\rho_{i=1}(\vec{r})$ and use it to construct the new $\phi_{i=1}(\vec{r})$ and iterate until self-consistency is achieved. The self-consistence solution of the equation gives $\rho^{\frac{1}{2}}(\vec{r})$ as the eigenfunction and $\mu$ as the eigenvalue. The corresponding $\phi(\vec{r})$ is obtained from $\rho^{\frac{1}{2}}(\vec{r})$ using Eq. (\ref{eq21}). This calculation is performed for different $a$ and $b$; the correct solution corresponds to those values of $a$ and $b$ that give the minimum $\mu$. 
\par
The self-consistent procedure outlined above is mathematically sound because $\rho^{\frac{1}{2}}(\vec{r})$ is always positive and therefore is like the ground-state wavefunction of a single particle with the LPS equation being the Schr$\ddot{o}$dinger equation for it. Hence the minimum value of the eigenenergy, which in this case is the chemical potential, represents the correct solution by variational principal.
\par 
The results of our calculations for the He-isoelectronic series are shown in Table \ref{tb2}.  It is clear from the Table that the self-consistently determined chemical potential is very close to the experimental chemical potential \cite{Lide}. Furthermore, its value is more negative than ($E_2-E_1$) obtained from the wavefunction calculations also shown in the Table \ref{tb1}. This is because in solving the LPS equation, $\mu$ is being minimized directly rather than being calculated the energy $E_2$ determined from the wavefunction. Thus the values obtained should be lower than $E_2-E_1$ as they indeed are. Because it is $\mu$ which is being minimized, the parameters $a$ and $b$ also differ slightly from those for the wavefunction calculations where $E_2$ is minimized \cite{Sech, Rabi}. If the wavefunction were exact the two would of course be the same. Next in Fig. (\ref{fig4}), we show the difference between the radial density as obtained by solving the LPS equation and the UG-density. As is clear from the figure the maximum difference between the two is of the order of $10^{-4}$.
\begingroup
\squeezetable
\begin{table}[H]
\caption{The optimized variational parameters $a$ and $b$, and the corresponding chemical potential $\mu$ for He-isoelectronic series as obtained by solving the LPS equation self-consistently. Experimental chemical potential $\mu_{expt}$ is also shown.\label{tb2}}
\begin{ruledtabular}
\begin{tabular}{lcccc}
 atom/ion               &  a   &   b    & $-\mu$ &    $-\mu_{expt}$\cite{Lide} \\  \hline
 H\textsuperscript{$-$} & 0.58 & 0.01   & 0.0273 & 0.0277            \\
 He                     & 0.90 & 0.13   & 0.9048 & 0.9036            \\
 Li\textsuperscript{+}  & 1.18 & 0.25   & 2.7813 & 2.7798            \\
 Be\textsuperscript{2+} & 1.43 & 0.38   & 5.6571 & 5.6557            \\
 B\textsuperscript{3+}  & 1.67 & 0.51   & 9.5326 & 9.5320            \\
 C\textsuperscript{4+}  & 1.90 & 0.64   &14.4080 & 14.4086           \\
 N\textsuperscript{5+}  & 2.11 & 0.77   &20.2833 & 20.2878           \\
 O\textsuperscript{6+}  & 2.32 & 0.90   &27.1585 & 27.1678           \\
 F\textsuperscript{7+}  & 2.51 & 1.03   &35.0336 & 35.0548           \\
 Ne\textsuperscript{8+} & 2.71 & 1.16   &43.9088 & 43.9449  
\end{tabular}
\end{ruledtabular}
\end{table}
\endgroup
\begin{figure}[h]
\caption{ \label{fig4} The difference between the density $r^2\rho_{LPS}^{SC}(r)$ obtained by solving the LPS equation self-consistently with the exact density $r^2\rho_{UG}(r)$ is shown.}
\includegraphics[width=0.5\textwidth]{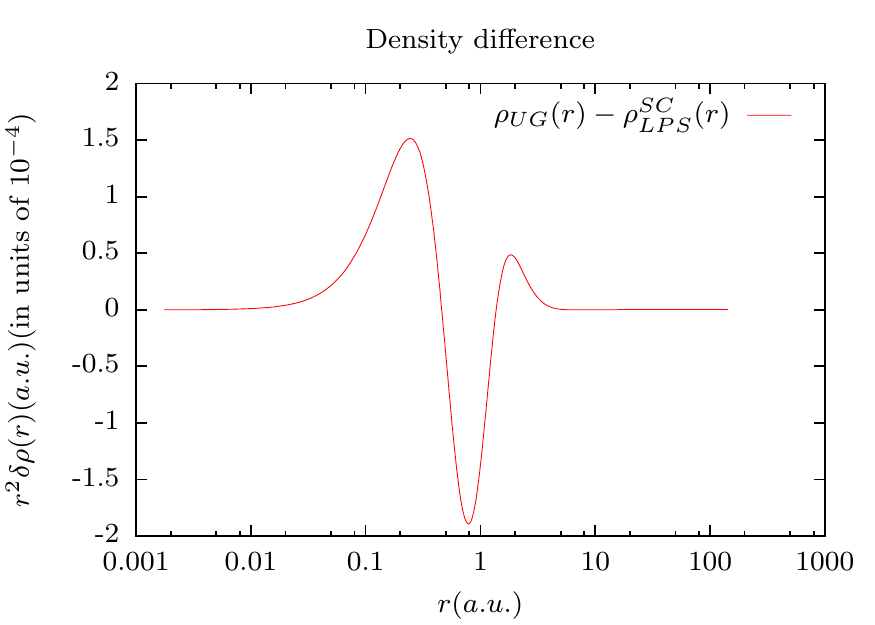}
\end{figure}

\par 
Finally we note that numerically, solving the LPS equation was simpler and more stable than solving the equation for $\phi(\vec{r})$ as was done in refs. \cite{Rabi, Baber}.
\section{Concluding remarks}
Using the Le Sech and modified Le Sech wavefunctions, we have constructed the effective potential $v_{eff}(\vec{r})$ of the LPS equation for two electron atomic systems. The solution of the LPS equation employing this $v_{eff}(\vec{r})$ leads to accurate densities and chemical potential for these systems. Encouraged by this, we have also solved the LPS equation self-consistently and shown that (with perfectly chosen correlated factor in the wavefunction) the resulting densities and chemical potentials are highly accurate. Further exploration in this direction for larger systems will be done in the future.  
   

\end{document}